\documentclass[sigconf]{acmart}
\AtBeginDocument{%
  }

\acmYear{2025}\copyrightyear{2025}
\setcopyright{acmlicensed}
\acmConference[Middleware '25]{26th ACM Middleware Conference}{December 15--19, 2025}{Nashville, TN, USA}
\acmBooktitle{26th ACM Middleware Conference (Middleware '25), December 15--19, 2025, Nashville, TN, USA}
\acmDOI{10.1145/3721462.3730950}
\acmISBN{979-8-4007-1554-9/25/12}

\acmSubmissionID{16}

\usepackage[]{hyperref}

\usepackage{algorithm}
\usepackage[noend]{algorithmic}

\usepackage{enumitem}

\usepackage{pifont}
\newcommand{\BlackCircled}[1]{\ding{\numexpr#1+201}}
\newcommand{\WhiteCircled}[1]{\ding{\numexpr#1+191}}

\newcommand{\bb}{{FreeRide}}
\newcommand{\pp}{pipeline parallelism}

\begin{document}

\title{\bb{}: Harvesting Bubbles in Pipeline Parallelism}

\author{Jiashu Zhang}
\affiliation{%
  \institution{University of Waterloo}
  \city{Waterloo}
  \country{Canada}}
\email{jiashu.zhang@uwaterloo.ca}

\author{Zihan Pan}
\affiliation{%
  \institution{University of Waterloo}
  \city{Waterloo}
  \country{Canada}}
\email{z82pan@uwaterloo.ca}

\author{Molly (Yiming) Xu}
\affiliation{%
  \institution{University of Waterloo}
  \city{Waterloo}
  \country{Canada}}
\email{molly.xu@uwaterloo.ca}

\author{Khuzaima Daudjee}
\affiliation{%
  \institution{University of Waterloo}
  \city{Waterloo}
  \country{Canada}}
\email{khuzaima.daudjee@uwaterloo.ca}

\author{Sihang Liu}
\affiliation{%
  \institution{University of Waterloo}
  \city{Waterloo}
  \country{Canada}}
\email{sihangliu@uwaterloo.ca}



\begin{abstract}
The occurrence of bubbles in pipeline parallelism is an inherent limitation that can account for more than 40\% of the large language model (LLM) long training times and is one of the main reasons for the under-utilization of GPU resources in LLM training.
Harvesting these bubbles for GPU side tasks can increase resource utilization and reduce training costs but comes with challenges. 
First, because bubbles are discontinuous with various shapes, programming side tasks becomes difficult while requiring excessive engineering effort.
Second, a side task can compete with pipeline training for GPU resources and incur significant overhead.
To address these challenges, we propose FreeRide, a middleware system that harvests the hard-to-utilize bubbles in pipeline parallelism systems to run generic GPU side tasks. 
FreeRide provides programmers with interfaces to implement side tasks easily, manages bubbles and side tasks during pipeline training, and controls access to GPU resources by side tasks to reduce overhead.
We demonstrate that FreeRide achieves almost 8\% average cost savings with a negligible overhead of about 1\% in training LLMs while serving model training, graph analytics, and image processing side tasks.
\end{abstract}

\maketitle 
\pagestyle{plain} 

\section{Introduction}

Large language models (LLMs) are usually trained on GPUs.
As these models continue to increase in size, their GPU memory requirements can easily outstrip the capacity of a single GPU \cite{zhang_opt_2022, villalobos_machine_2022}.
Consequently, to accommodate this increase in size and to boost the performance of pipeline training, it is a common practice to parallelize the training of LLMs across multiple GPUs distributed over several servers.

Pipeline parallelism is a prevalent training paradigm for LLMs using multiple GPUs.
In this paradigm, the model is divided into multiple stages, each consisting of several consecutive layers.
These stages are distributed across different GPUs.
During each training epoch, a batch of input data is split into multiple micro-batches.
Each micro-batch undergoes a forward propagation (FP) and a backward propagation (BP).
The FP and BP operations on different micro-batches are carried out in parallel by the pipeline training system at each stage.
The pipeline training system schedules these operations in each epoch to train LLMs~\cite{rasley_deepspeed_2020, kim_torchgpipe_2020, huang_gpipe_2019, fan_dapple_2021, liu_hanayo_2023, li_chimera_2021, harlap_pipedream_2018, narayanan_memory-efficient_2021, deepspeed_developers_pipeline_2023}.

An inherent limitation of \pp{} is \textit{bubbles} --- periods in pipeline training where the GPU stays idle due to unsatisfied dependencies between FP and BP operations~\cite{liu_hanayo_2023, li_chimera_2021}.
Experimentally, we observe that bubbles can constitute 42.4\% of the pipeline execution time, which results in significant under-utilization of GPU resources used to accelerate pipeline training.
Similar levels of under-utilization have also been reported in other studies~\cite{zhang_opt_2022, chowdhery_palm_2024}.

GPUs are crucial resources, especially those high-end models required for training LLMs~\cite{zhang_opt_2022, griffith_2023_the, strati_2024_ML}.
To enhance utilization, prior work has explored interleaving FP and BP operations~\cite{huang_gpipe_2019, harlap_pipedream_2018, narayanan_memory-efficient_2021, fan_dapple_2021}.
There have also been proposals to shard models into more stages and to deploy these stages on GPUs to better overlap the computation and communication \cite{li_chimera_2021, liu_hanayo_2023}.
These approaches are effective for intra-epoch bubbles because they change how operations are interleaved within a pipeline epoch.
However, they do not remove the inter-epoch bubbles that occur before and after a pipeline epoch.
Prior work has also proposed to decouple the computation of gradients for the input and model weights to mitigate inter-epoch bubbles~\cite{qi_zero_2023, tang_adaptive_2024}.
However, they increase the size of activations, exacerbating GPU memory consumption, a common bottleneck in training LLMs.

Given the difficulty and overhead incurred in eliminating these bubbles
, an alternative approach is to acknowledge their existence and utilize them by running additional workloads on a GPU.
For example, Bamboo~\cite{thorpe_bamboo_2023} uses bubbles to perform redundant computation for the successive layers to improve the reliability of pipeline training on spot instances.
PipeFisher computes second-order optimization based on the Fisher information matrix to increase the convergence speed of LLM training \cite{osawa_pipefisher_2023}.
However, Bamboo and PipeFisher only target specialized procedures that are tightly coupled with pipeline training, requiring the training system and the procedures to be highly customized.
Consequently, their approaches cannot be used for generic GPU workloads.

In this paper, we present \bb{}, a middleware system that bridges the gap between the available yet hard-to-utilize bubbles in \pp{} and the extra GPU workloads we run to harvest them.
We refer to these extra GPU workloads as \emph{side tasks}.
There are two main challenges that \bb{} has to overcome.
The first challenge is the programming complexity.
Bubbles are of various \emph{shapes}, i.e., their duration and available GPU memory.
Customizing side tasks for these bubbles by doing ad-hoc implementation requires enormous programming effort.
Second, LLM training requires high-end GPUs that are expensive and in high demand.
If side tasks interfere with the main pipeline training workload, e.g., overlapping their GPU execution with pipeline training or accessing more GPU resources than bubbles can provide, they will slow down pipeline training and increase training costs.

Our approach to overcoming the programming complexity is based on the observation that many GPU workloads naturally consist of small, repetitive steps, such as the epochs in model training that repeatedly load data and update model weights.
\bb{} operates between the pipeline parallel training and the generic GPU side tasks implemented by the user.
To reduce the programming effort, \bb{} introduces a framework that abstracts away the implementation details of side tasks, allowing programmers to adapt various side tasks to fit into the bubbles.
The key idea is to represent the life cycle of a side task, from its process creation to termination, as states in a state machine.
\bb{} provides two sets of unified interfaces --- the iterative interface that features lower performance overhead, and the imperative interface that features better versatility.
They facilitate the implementation of side tasks as state transitions with little engineering effort.
\bb{} manages side tasks through these interfaces and serves them during bubbles.

\bb{} limits the GPU resource consumption of side tasks through the automated side task profiler and the side task manager.
The side task profiler first captures the key performance characteristics of the newly implemented side tasks.
The side task manager coordinates a group of side task workers, one for each GPU in the platform, and assigns each of the side tasks to one of the workers based on the characteristics.
During pipeline training, bubbles are reported to the side task manager from the \bb{}-instrumented pipeline training system.
The side task manager starts side tasks when the bubble period begins and pauses them when the bubble ends.
A side task worker deploys each side task on top of CUDA MPS, which enables the concurrent execution of CUDA kernels from different processes~\cite{nvidia_developers_multi-process_2024} to limit the side task's GPU memory consumption and uses a containerized environment, e.g., Docker~\cite{bernstein_containers_2014} for isolation.
These components work collaboratively to serve side tasks during bubbles, achieving a low performance overhead on the primary pipeline training workload.

In summary, \bb{} is a middleware system that bridges the gap between the resourceful yet hard-to-utilize bubbles in \pp{} and the extra GPU workloads we run to harvest the bubbles.
It provides a holistic solution to manage and serve side tasks by leveraging bubbles in pipeline training, while maintaining minimal performance overhead and requiring low programming effort. 
We evaluate \bb{} by deploying it to run side tasks alongside DeepSpeed that runs pipeline training~\cite{rasley_deepspeed_2020}.
We measure the time increase in pipeline training as the performance overhead caused by side tasks.
As the throughput of different side tasks is not directly comparable with the pipeline training workload, we use the cost of GPUs as a unified metric, i.e., the cost of the extra execution time from co-locating side tasks with pipeline training vs. the cost saved from running side tasks that otherwise would run on dedicated lower-tier GPUs.

The contributions of this paper are as follows:
\begin{itemize}[leftmargin=*]
\item We study the bubbles in pipeline parallelism, present their various shapes in terms of duration and available GPU memory, and demonstrate their potential for side tasks.
\item We present the programming framework and the interfaces of \bb{}\footnote{\url{https://github.com/jiashu-z/freeride}} based on a state machine abstraction to implement generic side tasks with little engineering effort.
\item We evaluate \bb{} with model training, graph analytics, and image processing side tasks to demonstrate \bb{}'s effectiveness in harvesting bubbles in \pp{} while reducing performance overhead.
\item By serving side tasks based on the iterative interface, FreeRide achieves average \emph{cost savings} of 7.8\% with a low performance overhead (time increase in pipeline parallel training) of 1.1\%. This is significantly better than using CUDA MPS~\cite{nvidia_developers_multi-process_2024} directly to co-locate the tasks, which results in a 4.5\% \emph{cost increase} and 48.7\% overhead. When handling a mix of these three types of side tasks, FreeRide achieves 10.1\% cost savings with a 1.1\% overhead.

\end{itemize}

\section{Background and Motivation}
\label{sec:background_and_motivation}

In this section, we provide an overview of pipeline parallelism for training LLMs, bubbles in the pipeline, and motivation for utilizing the bubbles to execute generic workloads. 

\subsection{Pipeline Parallelism and Bubbles}
\label{sec:pipeline-parallelism}

Pipeline parallelism is a widely used paradigm for distributed training of LLMs that far exceed the memory capacity of a single GPU~\cite{zhang_opt_2022, rasley_deepspeed_2020, smith_2022_using}.
In \pp{}, the model is divided into multiple stages, where each stage executes several consecutive layers of the model.
These stages are deployed across different GPUs to form a pipeline.
To parallelize the computation at each stage and reduce GPU memory consumption, one batch of input data is split into micro-batches during each training epoch.
Each micro-batch undergoes forward propagation (FP) and backward propagation (BP).
In both FP and BP operations, after a stage finishes processing one micro-batch of data, it passes its output to the next stage and immediately moves on to the next micro-batch.
The FP and BP operations constitute the epochs in pipeline training systems~\cite{rasley_deepspeed_2020, deepspeed_developers_pipeline_2023, kim_torchgpipe_2020, huang_gpipe_2019, fan_dapple_2021, liu_hanayo_2023, li_chimera_2021, harlap_pipedream_2018, narayanan_memory-efficient_2021}.
A myriad of frameworks have been developed to support pipeline training.
For example, DeepSpeed~\cite{rasley_deepspeed_2020} and Megatron~\cite{shoeybi_2020_megatronlm} are extensively used to train various LLMs such as OPT~\cite{zhang_opt_2022}, Turing-NLG~\cite{hagen_turing-nlg_2020}, and MT-NLG~\cite{smith_2022_using}.

\begin{figure}
    \centering
    \includegraphics[width=\columnwidth]{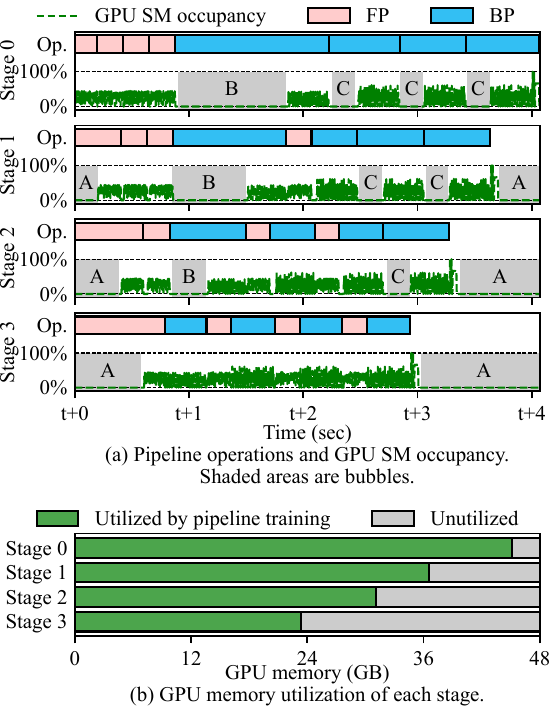}
    \caption{A pipeline training epoch in DeepSpeed.}
    \label{fig:bubble_study}
    \Description[]{}
\end{figure}

There are periods in pipeline training when the GPU streaming multiprocessor (SM) occupancy is low, as depicted by the green curves in Figure~\ref{fig:bubble_study}(a).
We refer to these periods as {\em bubbles} in the pipeline, which are marked as shaded areas.
Bubbles inherently exist in \pp{} and occur repetitively throughout training, as they are fundamentally caused by unsatisfied dependencies between FP and BP operations~\cite{li_chimera_2021, liu_hanayo_2023}.
In the example of Figure~\ref{fig:bubble_study}, Stage~1 must wait for input from Stage 0 before it can execute its first FP operation, creating a bubble in Stage 1 that starts from $t+0$.

\subsection{Bubble Characterization}
\label{sec:bubble-study}

To study bubbles in \pp{}, we train a 3.6B-parameter LLM adapted from previous work~\cite{choi_envpipe_2023, karpathy_nanogpt_2024} using DeepSpeed~\cite{rasley_deepspeed_2020} on a 4-GPU server (detailed setup in Section~\ref{sec:meth}).
The training is deployed as a 4-stage pipeline, and each stage takes one GPU as shown in Figure~\ref{fig:bubble_study}.
Overall, we observe that bubbles exhibit different characteristics across all 4 stages. 
Next, we take a closer look at each type of bubble. 

\subsubsection{Bubble Categorization}
We categorize bubbles into 3 types based on their positions in the training pipeline and their causes.

\noindent$\bullet$~\textbf{Type-A bubbles} appear at the start and end of each epoch in all stages except for the first stage.
They are due to cascading dependencies between operations at the start and end of an epoch.
When an epoch starts, the FP operations start at Stage~0, while all other stages wait for input data from their preceding stages to start their first FP operation. 
Likewise, at the end of an epoch, the last BP operation starts at Stage~3 and all other stages wait for their succeeding stages to start their last BP operation.

\noindent$\bullet$~\textbf{Type-B bubbles} occur in the middle of each epoch on all stages except the last one.
They are caused by dependencies between interleaved FP and BP operations.
Once the first FP operation reaches the last stage, all previous stages must wait for the corresponding BP operation before they can proceed with other operations, which causes Type-B bubbles.

\noindent$\bullet$~\textbf{Type-C bubbles} also occur in the middle of each epoch.
Since BP operations typically take longer than FP operations~\cite{zheng_alpa_2022}, interleaved yet unaligned FP and BP operations create bubbles in each stage except the last.
For instance, in Figure~\ref{fig:bubble_study}(a), when Stage~2 finishes its third BP operation, it must wait for input to its fourth BP operation, which is still being processed in Stage~3, causing a type-C bubble.

\noindent{\textbf{Duration.}}
In our training setup, the duration of a bubble ranges from 0.22 to 1.04 seconds, depending on its type and stage.
The duration increases for Type-A bubbles but decreases for Type-B bubbles from Stage~0 to Stage~3.
This is because of the cascading dependency from Stage~3 to Stage~0 for Type-A bubbles and from Stage~0 to Stage~3 for Type-B bubbles.
For example, a Type-B bubble at Stage~2 is due to an unfinished BP operation at Stage~3, with the same bubble at Stage~1 caused by Stage~2.
The accumulated time to satisfy dependencies elongates Type-A or Type-B bubbles at later stages.
However, Type-C bubbles are caused by unaligned FP and BP operations.
Therefore, they have a short duration and do not exhibit the same stage-dependent variations.

\noindent{\textbf{Available GPU Memory.}}
Determined by the stage, the available GPU memory of a bubble ranges from less than 3 GB to more than 20 GB in our setup.
As shown by Figure~\ref{fig:bubble_study}(b), within a stage, the GPU memory consumption of pipeline training remains the same.
Thus, the bubbles within the same stage have the same amount of available GPU memory.
Because the later stages consume less GPU memory to store activations used by BP operations~\cite{liu_hanayo_2023}, the available GPU memory increases from Stage 0 to Stage 3.

\begin{figure}
    \centering
    \includegraphics[width=\columnwidth]{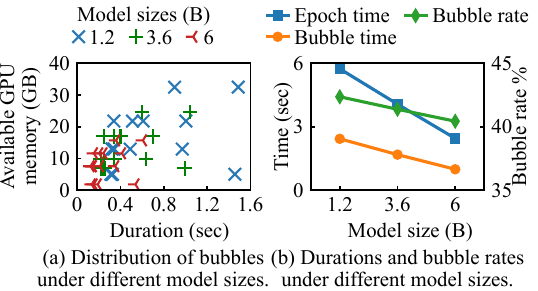}
    \caption{Statistics of bubbles under different model sizes.}
    \label{fig:duration-memory}
    \Description[]{}
\end{figure}

We further study pipeline training of models of different sizes.
As shown in Figure~\ref{fig:duration-memory}(a), bubble shapes differ. 
Overall, bubbles in larger LLMs have less available memory and shorter duration, but the distributions are similar as training follows the same pipeline schedule. 
Even larger models do not eliminate the inherently exist bubbles. 
Under the same configuration, bubbles have the same characteristics during training, as epochs are repetitive and stable.

\subsubsection{Bubble Rate}

Besides the bubble shape, we evaluate the overall \emph{bubble rate}, i.e., the total bubble time over pipeline training time.
When the model size increases from 1.2B to 6B parameters, as shown in Figure~\ref{fig:duration-memory}(b), both the per-epoch time in pipeline training and the total per-stage bubble time decrease.
Therefore, the bubble rate drops only slightly from 42.4\% to 40.4\%.
We also evaluate a larger micro-batch number, i.e., an increase from 4 (used in Figures~\ref{fig:bubble_study} and \ref{fig:duration-memory}) to 8.
The bubble rate drops to 26.2\% as each epoch takes longer.

Prior work has focused on reducing bubbles in pipeline parallelism.
One approach is designing different ways of interleaving FP and BP operations~\cite{huang_gpipe_2019, harlap_pipedream_2018, narayanan_memory-efficient_2021, fan_dapple_2021}; another type of optimization divides the model into more stages and orchestrates their deployment to overlap computation and communication~\cite{li_chimera_2021, liu_hanayo_2023}.
These approaches optimize the scheduling strategies and interleave FP and BP operations within an epoch.
Therefore, they are effective for Type-B and Type-C bubbles that appear inside an epoch but not for Type-A bubbles. 
There has also been work that reduces Type-A bubbles by decoupling the computation of gradients for the input and the model weights~\cite{qi_zero_2023, tang_adaptive_2024}.  This comes at a cost of higher GPU memory usage due to the extra activation storage, exacerbating the GPU memory bottleneck in LLM training.
Despite these efforts, none of the approaches fully eliminate bubbles in pipeline training.

\subsection{Utilizing Bubbles}
\label{sec:possibility}

The difficulties in mitigating bubbles in \pp{} motivate an alternative approach --- acknowledging their existence and leveraging their resources by allocating additional GPU workloads to them.
Prior work has utilized bubbles to run procedures that enhance pipeline training.
For example, Bamboo uses bubbles to perform redundant computation for successive layers to improve the reliability of pipeline training on spot instances~\cite{thorpe_bamboo_2023}; PipeFisher computes second-order optimization based on the Fisher information matrix to speed up convergence~\cite{osawa_pipefisher_2023}.
However, they tightly couple the pipeline training system with specialized procedures.
Implementing specialized procedures is complicated, especially since such customization should consider various bubble shapes --- with durations ranging from 0.22 to 1.04 seconds, and available GPU memory from less than 3 GB to more than 20 GB on each GPU (Section~\ref{sec:bubble-study}).

GPUs used for training are generally compute-rich, with sufficient GPU memory available during the bubbles to accommodate other GPU workloads. 
Therefore, bubbles can be used to run workloads that otherwise require dedicated GPUs. 
For instance, training a ResNet18 model with batch size 64 takes only 2.63 GB of GPU memory with each iteration taking only 30.4 ms on our platform --- small enough to fit into most of the bubbles in Figure~\ref{fig:bubble_study}(a).
By doing so, the resources available in bubbles present prime opportunities for serving GPU workloads, which can amortize the expensive cost of LLM training with effective GPU workload execution.
We refer to these GPU workloads served during bubbles as \emph{side tasks}.
Prior solutions that target specialized co-running procedures \cite{osawa_pipefisher_2023, thorpe_bamboo_2023} do not apply to generic workloads. 

In this work, we aim to \emph{make bubble resources available to generic workloads, allowing for a programmable and efficient use of bubbles.}

\subsection{Challenges}
\label{sec:challenge}

To execute generic GPU side tasks during bubbles, we identify two major challenges.

\textbf{Challenge 1: programming effort required to implement side tasks.}
Typically, GPU workloads are implemented assuming that they have access to the full GPU and can run continuously until they finish execution.
However, bubbles are \emph{intermittent} and largely vary in duration, as discussed in Section~\ref{sec:bubble-study}. 
A side task should be tailored to bubble patterns --- the side task automatically pauses or resumes when a bubble ends or starts.
Customizing the training framework to embed side tasks is conceptually feasible but limits the flexibility of implementing and executing generic GPU workloads, much like the limitations from prior work on co-running specialized procedures~\cite{osawa_pipefisher_2023, thorpe_bamboo_2023}.

\textbf{Challenge 2: limiting the impact of side tasks.}
LLM training can span months on expensive high-end GPUs and cost millions of dollars~\cite{zhang_opt_2022, li_openais_2020}.
Even with side tasks placed in the under-utilized bubbles, they may still interfere with pipeline training, significantly increasing the cost of LLM training and offsetting the benefit of running side tasks.
However, limiting the impact of side tasks is not trivial.
As the shape of bubbles varies, naively implementing side tasks may consume more resources than bubbles have --- exceeding the duration of bubbles or even crashing the main task due to excessive GPU memory allocation.
Ideally, bubbles should be utilized without impacting the more expensive and prioritized LLM training task.

\section{\bb{} Design Overview}
\label{sec:high_level_idea}

\bb{} is our middleware system that addresses the aforementioned challenges in utilizing bubbles in pipeline training to serve generic GPU side tasks.
It minimizes the performance impact of side tasks on pipeline training.
In this section, we present the high-level ideas of \bb{}.

\subsection{Side Task Programming Interface}

Given the high cost and priority of the main pipeline training workload,
the side task should not overlap with this main task to avoid competing for GPU resources. 
This requirement is challenging from a programmer's perspective, as it is difficult to tailor every workload to different bubble shapes.
We observe that GPU workloads are not monolithic, rather, they can be often divided into small, repeated \emph{steps} with largely predictable per-step duration and resource consumption, i.e., GPU memory.
For example, epochs in model training, iterations in graph analytical workloads~\cite{xu_gardenia_2019,page_pagerank_1999,koren_matrix_2009}, and steps to process each image in image-processing workloads~\cite{nvidia_developers_image_2019} all follow this pattern. 
On the other hand, bubbles also demonstrate repeating and predictable patterns, as discussed in Section~\ref{sec:bubble-study}.

With these observations in mind, our idea is to provide an iterative programming interface that can incorporate the step-by-step execution of side tasks to bubbles with various shapes.
The user only has to implement the side task step without being concerned with the bubble shapes, and the bubbles can serve these side tasks with largely predictable durations to avoid lack of GPU resource or overlapping of side task execution and pipeline training.
The iterative programming interface provided by \bb{} employs a state machine abstraction for the life cycle of a side task composed of different states during its execution.
The execution of side tasks within bubbles can be implemented as state transition functions in Figure~\ref{fig:programming-model} (details in Section~\ref{sec:programming-model}).
\bb{} works as the middleware layer in between the side tasks and the bubbles of pipeline training, managing the side tasks' start and pause through controlling their state transitions.
In this way, \bb{} fits the side tasks into bubbles and minimizes the performance impact on pipeline training.

We recognize that not all GPU workloads can be easily adapted to our iterative model. 
To accommodate these workloads, \bb{} provides the imperative interface as an alternative.
The idea is to enable pausing and resuming of execution using transparent signals from a lower level.
For this reason, it does not require complex adaption but comes with a slightly higher performance overhead.
We discuss both interfaces in Section~\ref{sec:template}.

\subsection{Profiling-guided Side Task Management} 

As bubbles have different shapes, when a side task is newly added to \bb{}, it should be assigned to a stage whose bubbles have enough GPU memory available.
When a side task is served during bubbles, there should be mechanisms that make sure the side task does not consume more resources than available by the bubbles to minimize the overhead of \bb{}, e.g., excessively allocating GPU memory or not pausing when a bubble ends.

\begin{figure}
    \centering
    \includegraphics[width=\columnwidth]{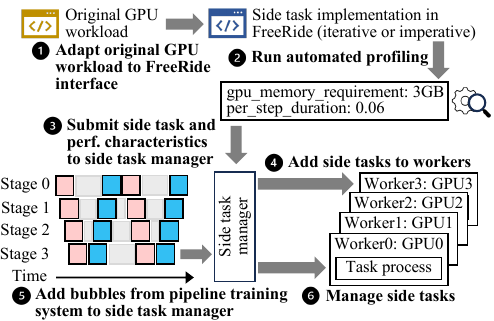}
    \caption{Workflow of \bb{}.}
    \label{fig:overview}
    \Description[]{}
\end{figure}

To judiciously manage side tasks on bubbles, \bb{} leverages profiling to understand the shapes of bubbles, which can be done offline or during the first few epochs of pipeline training.
Then, when a side task is newly submitted to \bb{}, as shown in Figure~\ref{fig:overview}, \bb{}'s automated side task profiler tracks its GPU memory consumption and per-step duration.
During execution time, \bb{} employs one \textit{side task manager} and multiple \textit{side task workers}, one for each GPU.
The side task manager assigns the newly submitted side task to one of the side task workers, which will create the side task process, based on the resulting profile.
We instrument DeepSpeed to report the start timestamps and duration of bubbles to the side task manager that will initiate state transitions of each side task through remote procedure calls (RPCs) at the start and end of each bubble.

\bb{} minimizes performance overhead on the main pipeline training workload by limiting the GPU resource consumed by side tasks (Section~\ref{sec:resource-control}).
For GPU memory, the side task worker of \bb{} leverages CUDA MPS~\cite{nvidia_developers_multi-process_2024} to impose a limit on GPU memory consumed by a side task process.
For GPU execution time, \bb{} uses a twofold mechanism --- 
a \emph{program-directed} mechanism through the programming interface, and a \emph{framework-enforced} mechanism based on the side task manager and workers.
In addition, the side task worker can deploy side task processes in Docker containers~\cite{bernstein_containers_2014} for isolation.

\subsection{\bb{} Workflow}
Putting the aforementioned ideas together, we present the workflow of \bb{} in Figure~\ref{fig:overview}.
First, programmers adapt their side task implementation using the interface provided by \bb{} (step~\BlackCircled{1}).
\bb{} then automatically generates a profile of the side task's characteristics (step~\BlackCircled{2}), which is submitted with the side task to the side task manager of \bb{} (step~\BlackCircled{3}).
After the side task is submitted, based on the memory allocation of pipeline training and the characteristics of the side task, the side task manager will assign this side task to one of the workers (step~\BlackCircled{4}).
When the main pipeline training workload is running, 
the side task manager continuously adds bubbles from the instrumented LLM training framework (step~\BlackCircled{5}); at the same time, it starts/pauses side tasks based on the available bubbles (step~\BlackCircled{6}).

\begin{figure}
    \centering
    \includegraphics[width=\columnwidth]{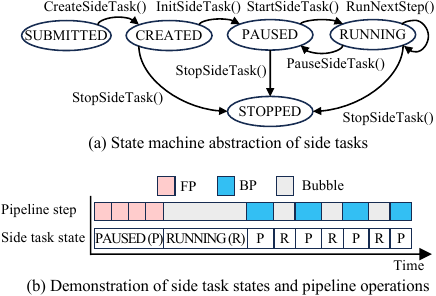}
    \caption{State transitions in a side task program.}
    \label{fig:programming-model}
    \Description[]{}
\end{figure}

\section{Implementation of \bb{}}
\label{sec:freeride}

In this section, we first introduce how \bb{} supports side tasks through its framework and interfaces.
Then, we present details of \bb{}'s profiling-guided side task management.
Finally, we discuss \bb{}'s GPU resource limit mechanisms including the implementation details.

\subsection{Programming Framework of \bb{}}
\label{sec:programming-model}

Figure~\ref{fig:programming-model}(a) describes the programming framework of a side task.
The framework's core is a state machine with five states and six state transitions.
These five states capture the life cycle of a side task, from process creation to process termination, and correspond to different usage of hardware resources, e.g., GPU memory and GPU execution time.
The six state transitions are used by the programmer to implement the user-defined logic of a side task.
The programmer can override the state transition functions to customize their behavior, e.g., allocating or releasing hardware resources or performing computation on GPU.
Once the side task is implemented, \bb{} automatically handles the state transitions at runtime. 
Next, we discuss the states and their transitions.

$\bullet$~\texttt{SUBMITTED}. This state means that \bb{} has profiled a task and submitted it to the side task manager, but the side task worker has not created the side task process yet.
State transition \texttt{CreateSideTask()} happens automatically after the side task manager assigns a side task to a worker and the worker creates the side task process.

$\bullet$~\texttt{CREATED}.
In this state, the worker has created the side task process, and this process has loaded its context to the main memory but not to the GPU memory.
Take a model training side task as an example.
When it is in the \texttt{CREATED} state, it has already created and initialized the dataset, the data loader, the loss function, and the optimizer states in CPU memory.
However, the side task process will not load them into GPU memory until the side task manager initiates the state transition \texttt{InitSideTask()}.
The state transition \texttt{InitSideTask()}, initiated by the side task manager, means that the side task will finish initialization.

$\bullet$~\texttt{PAUSED}.
This state is where the side task starts to use GPU memory.
The side task process has loaded its context, e.g., model weights and optimizer states, in the GPU memory.
However, this process waits in the \texttt{PAUSED} state until the side task manager transitions its state to \texttt{RUNNING} through \texttt{StartSideTask()}.

$\bullet$~\texttt{RUNNING}.
In this state, the side task executes the step-wise GPU workload.
Referring to the example above of the model training side task, this step involves reading the next batch, computing the output and loss, updating the model weights, and resetting the optimizer states.
The side task iteratively enters the \texttt{RunNextStep()} state transition to execute these steps until the side task manager transitions its state through \texttt{PauseSideTask()}.
Therefore, in this state, the side task process uses both GPU memory and SMs.

$\bullet$~\texttt{STOPPED}.
This state marks the end of the life cycle of a side task, where the side task process releases all of its hardware resources and terminates.
It can be transitioned from states \texttt{CREATED}, \texttt{PAUSED}, and \texttt{RUNNING} through \texttt{StopSideTask()} initiated by the side task manager.

Figure~\ref{fig:programming-model}(b) shows state transitions of a side task in Stage~0 of Figure~\ref{fig:bubble_study}.
Initially, the side task is in the \texttt{PAUSED} (\texttt{P}) state.
After four FP operations in the main training workload have finished, a bubble starts and the side task manager initiates \texttt{StartSideTask()} to transition the side task to the \texttt{RUNNING} (\texttt{R}) state.
After the first bubble ends, the side task manager initiates \texttt{PauseSideTask()} to pause the side task.
Then, the main training workload has BP operations and bubbles interleaved, leading to back-and-forth transitions between  \texttt{PAUSED} and \texttt{RUNNING} states of the side task.

\subsection{Interface for Side Task Implementation}
\label{sec:template}
Given the \bb{} programming framework, the next step is to implement side tasks, which have two requirements.
First, the programmer should be able to implement the side task in a way that can pause at the end of a bubble and resume at the start of the next bubble.
Second, the side task should be able to communicate with the side task manager to receive state transition RPCs (Section~\ref{sec:instrumentation}) for pausing and resuming.
To lift programming burdens, \bb{} provides two programming interfaces supported in C++ and Python.
Once implemented using either interface, \bb{} will handle the side tasks and their state transitions transparently at runtime. 
We discuss both interfaces next.

\noindent{\textbf{Iterative programming interface.}}
This is the preferred interface for side tasks in \bb{}.
It periodically checks whether the side task manager has initiated any state transitions.
If so, it executes the state transition functions in Figure~\ref{fig:programming-model}(a) and updates the state of the side task.
Then, if the side task is currently in the \texttt{RUNNING} state, it executes \texttt{RunNextStep()}.
The programmer only has to override these transition functions to implement the side task.
Pausing and resuming the side task, the transition of states, and communication with the \bb{} side task manager are all handled by the interface itself.
GPU workloads that are naturally step-wise, e.g., model training, can be easily adapted to the iterative interface.
We will discuss the adaption to this interface in Section~\ref{sec:example} using an example.

\noindent{\textbf{Imperative programming interface.}}
Not all side tasks can be explicitly implemented step-wise.
Therefore, \bb{} provides the imperative interface as a fallback solution.
The core is the function \texttt{RunGpuWorkload()} that allows the programmer to implement generic GPU side tasks without breaking them into steps.
When the side task manager changes the state of the side task to \texttt{RUNNING} for the first time, the interface calls the \texttt{RunGpuWorkload()} function to execute the side task.
The interface implements the pausing and resuming through signals (\texttt{SIGTSTP} and \texttt{SIGCONT}~\cite{job_control_signals}) and calls \texttt{StartSideTask()} and \texttt{PauseSideTask()} inside the handlers of the two signals.
The imperative interface offers better versatility at the cost of higher performance overhead (discussed in Section~\ref{sec:example} and evaluated in Section~\ref{sec:end_to_end}).

\subsection{Profiling Bubbles and Side Tasks}
\label{sec:profiling}

\noindent\textbf{Bubbles.}
To know the shapes of bubbles before serving side tasks with them, \bb{} runs DeepSpeed, monitors its estimated SM occupancy and GPU memory consumption through the PyTorch profiler~\cite{pytorch_profiler}, and automatically measures each bubble's duration and available GPU memory.
Since the pipeline schedule determines bubbles, this offline profiling is done only once for each model and pipeline scheduling on the same hardware platform.

\noindent\textbf{Side tasks.}
After the programmer has implemented the side task, \bb{} profiles it with the automated profiling tool for its performance characteristics of GPU memory consumption and per-step duration, which \bb{} uses for side task management and limiting GPU resources.
For side tasks implemented using the iterative interface, this procedure is fully automated.
The profiling tool runs the side task, monitors its GPU memory consumption, and records the timestamps at the start and end of \texttt{RunNextStep()} state transition for the per-step duration.
For side tasks implemented using the imperative interface, the tool profiles GPU memory consumption in the same way.
However, since the side task is not step-wise, the automated profiling tool does not measure the per-step duration.

\subsection{Side Task Management}
\label{sec:scheduler}

\bb{}'s side task management has two main roles.
First, upon receiving a new side task, the side task manager assigns it to a suitable side task worker. 
Second, when the pipeline training system adds bubbles to the side task manager, the side task manager initiates the state transitions of side tasks (Figure~\ref{fig:programming-model}(a)) through RPCs.
In this way, the side tasks are only served during bubbles and do not compete for GPU resources with the main pipeline training workload.

To keep track of side tasks and workers, the side task manager maintains the following fields for each worker, used by Algorithms~\ref{alg:submit_side_task} and \ref{alg:schedule} for side task management:
\begin{itemize}[leftmargin=*]
    \item \textit{GPUMem}: the available GPU memory size.
    \item \textit{TaskQueue}: the queue of side tasks ordered by submission timestamps.
    \item \textit{CurrentTask}: the side task that is currently served.
    \item \textit{CurrentBubble}: the bubble that is currently valid.
\end{itemize}

\begin{algorithm}
\caption{Procedure upon a new side task.}
\label{alg:submit_side_task}
\begin{algorithmic}[1]
\STATE {\bfseries Input: } new task $Task$, workers' metadata $Workers$
\STATE $MinNumTasks \leftarrow \infty$
\STATE $SelectedWorker \leftarrow None$
\FORALL{$Worker$ {\bfseries in} $Workers$}
    \IF{$Worker.GPUMem > Task.GPUMem$}
        \STATE $NumTasks \leftarrow Worker.GetTaskNum()$
        \IF{$NumTasks < MinNumTasks$}
            \STATE $MinNumTasks \leftarrow NumTasks$
            \STATE $SelectedWorker \leftarrow Worker$
        \ENDIF
    \ENDIF
\ENDFOR
\IF{$SelectedWorker \neq None$}
    \STATE $SelectedWorker.Add(Task)$
\ELSE
    \STATE $RejectSideTask()$
\ENDIF
\end{algorithmic}
\end{algorithm}

Algorithm~\ref{alg:submit_side_task} describes how the side task manager assigns side tasks to workers.
When the side task manager receives a new side task together with its GPU memory requirement (through profiling, Section~\ref{sec:profiling}), it first filters out all workers with enough available GPU memory (lines 4---5).
Then, from these workers, it selects the one with the smallest number of tasks (lines 6---9).
If the side task manager has selected a worker, it will assign the side task to that worker (lines 10---11).
Otherwise, it will reject the side task because of insufficient GPU memory (line 13).

\begin{algorithm}
\caption{Managing bubbles and side tasks.}
\label{alg:schedule}
\begin{algorithmic}[1]
\STATE {\bfseries Input: } workers' metadata $Workers$
\WHILE{$SideTaskManagerIsRunning$}
    \FORALL{$Worker$ {\bfseries in} $Workers$}
        \IF{$Worker.CurrentBubble \neq None$}
            \IF{$Worker.CurrentBubble.HasEnded()$}
                \IF{$Worker.CurrentTask \neq None$}
                    \STATE $Worker.CurrentTask.PauseSideTask()$
                \ENDIF
                \STATE $Worker.CurrentBubble \leftarrow None$
            \ENDIF
        \ENDIF
        \IF{$Worker.HasNewBubble()$}
            \STATE $Worker.UpdateCurrentBubble()$
            \IF{$Worker.CurrentTask = None$}
                \IF{$Worker.TaskQueue.IsEmpty()$}
                    \STATE $continue$
                \ENDIF
                \STATE $nextTask \leftarrow Worker.TaskQueue.Next()$
                \STATE $Worker.CurrentTask \leftarrow nextTask$
            \ENDIF
            \IF{$Worker.CurrentTask.IsCreated()$}
                \STATE $Worker.CurrentTask.InitSideTask()$
            \ELSIF{$Worker.CurrentTask.IsPaused()$}
                \STATE $Worker.CurrentTask.StartSideTask()$
            \ENDIF
        \ENDIF
    \ENDFOR
\ENDWHILE
\end{algorithmic}
\end{algorithm}

Algorithm~\ref{alg:schedule} describes how the side task manager manages bubbles and side tasks during pipeline training.
The side task manager iterates through all workers (line 3).
If \textit{CurrentBubble} has just ended for a worker, the side task manager will pause \textit{CurrentTask} of the worker and clear \textit{CurrentBubble} (lines 4---8).
Upon a new bubble, the side task manager updates the \textit{CurrentBubble} of this worker (lines 9---10).
It then checks if the worker has a \textit{CurrentTask}.
If not, it will select the one with the smallest submission timestamp from \textit{TaskQueue} as \textit{CurrentTask} (lines 11--15).
After that, 
the side task manager initiates \texttt{InitSideTask()} if the newly added \textit{CurrentTask} is in \texttt{CREATED} state (lines 16---17); 
otherwise, its state is \texttt{PAUSED} and the side task manager initiates \texttt{StartSideTask()} (lines 18---19).

\subsection{GPU Resource Limit}
\label{sec:resource-control}

In this section, we introduce the mechanisms in \bb{} that reduce the impact of side tasks on the main pipeline training workload through side task resource control for both GPU memory and GPU execution time.

\noindent\textbf{GPU Memory.}
\bb{} leverages MPS to impose GPU memory limit on side tasks.
I.e., when a worker creates a side task, it sets GPU memory limits using MPS.
The side task process triggers an out-of-memory (OOM) error when its memory consumption exceeds the limit, but other processes remain unaffected.
However, \bb{} is also compatible with other mechanisms for limiting GPU memory, e.g., multi-instance GPU (MIG)~\cite{nvidia_developers_mig} or manually implemented accounting through intercepting CUDA kernel calls~\cite{strati_orion_2024}.

\noindent\textbf{GPU Execution Time.}
\bb{} limits GPU execution time using two mechanisms.
(1) \underline{\emph{The program-directed mechanism}} is tailored for the iterative interface.
When the side task manager makes an RPC to initiate \texttt{StartSideTask()} state transition of a side task, it also sends the end time of this bubble to the side task.
After the state transition finishes, the side task enters the \texttt{RUNNING} state.
Before the side task automatically starts \texttt{RunNextStep()}, the program-directed mechanism checks if the remaining time of the bubble is enough for the side task to execute the next step.
The side task will only execute the next step if the remaining time exceeds the per-step duration.
(2) \underline{\emph{The framework-enforced mechanism}} supports side tasks implemented using the imperative interface and is also a fallback mechanism for the iterative interface.
After the side task manager initiates the \texttt{PauseSideTask()} state transition for a side task, it waits for a short grace period before checking the \textit{last paused timestamp} --- a timestamp maintained by the interface that records the last time the side task was paused. 
If this timestamp is not updated after the state transition begins, the side task manager assumes that the interface failed to pause the side task correctly and subsequently instructs the corresponding worker to terminate the side task process using \texttt{SIGKILL}.
The side task initialization, \texttt{InitSideTask}, which runs only once throughout the life cycle of a side task, is also protected by this mechanism.

\subsection{Implementation}
\label{sec:instrumentation}
We use DeepSpeed 0.12.2~\cite{deepspeed_0_12_2} as the framework for pipeline training.
We modify DeepSpeed in three places with 55 lines of code:
(1) before the start and at the end of an epoch for Type-A bubbles,
(2) after all FP operations preceding the first BP operation for Type-B bubbles, and (3) after the first BP operation following the last FP operation for Type-C bubbles.
The instrumented code reports bubbles to the side task manager in \bb{}, as shown in step \BlackCircled{5} of Figure~\ref{fig:overview}.
The modifications are done once, as the framework can be used for training different models. 

\begin{figure}
    \centering
    \includegraphics[width=\columnwidth]{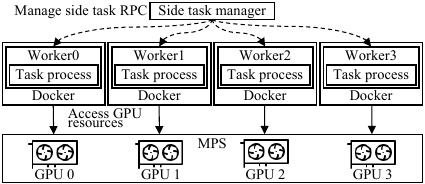}
    \caption{Architecture of \bb{}.}
    \label{fig:impl}
    \Description[]{}
\end{figure}

To isolate the side task processes from the pipeline training process, \bb{} deploys workers (and side tasks of these workers) inside Docker containers, as illustrated in Figure~\ref{fig:impl}.
\bb{} implements the side task manager and each side task worker in separate processes.
Communication among the pipeline training system, side tasks, and \bb{} components is facilitated through RPCs utilizing gRPC~\cite{grpc}.

\section{Use of Side Tasks Interface}
\label{sec:example}

\begin{figure*}
    \centering
    \includegraphics[width=0.8\textwidth]{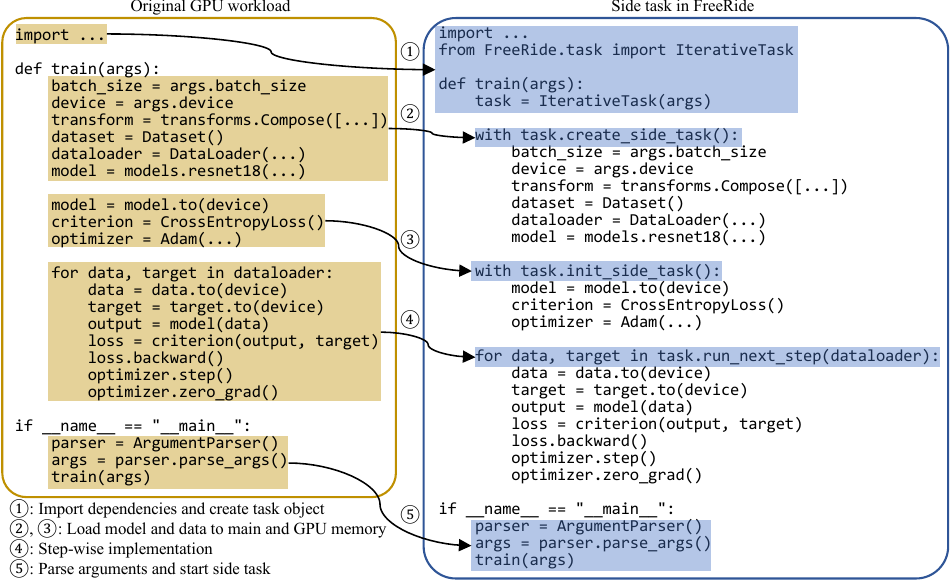}
    \caption{Example of implementing ResNet18 training using the iterative interface of \bb{}.}
    \label{fig:example}
    \Description[]{}
\end{figure*}

This section describes \bb{}'s iterative interface and imperative interface in detail.

\noindent{\textbf{Iterative programming interface.}}
Figure~\ref{fig:example} is an example of implementing a side task to train ResNet18 using the iterative interface of \bb{} in Python.
Less important lines such as importing dependencies and parsing arguments are simplified. 
Porting this example mainly involves 6 steps.
Step~\WhiteCircled{1}: import \bb{} dependencies and inherit the iterative interface class, which includes an implementation for the state machine abstraction, communication with the side task manager, and the program-directed mechanism to limit the GPU execution time.
The programmer only needs to adapt the original GPU workload to the interface.
Steps~\WhiteCircled{2} and \WhiteCircled{3}: implement the side task initialization in 2 state transition functions, \texttt{CreateSideTask()} and \texttt{InitSideTask()}, to load the context into main memory and GPU memory respectively.
Step~\WhiteCircled{4}: wrap the original loop implementation with \texttt{RunNextStep()}.
Step~\WhiteCircled{5}: the main function parses arguments and runs the side task interface.

Most modifications are trivial, e.g., wrapping implementations with side task state transition functions in Step \WhiteCircled{2}, \WhiteCircled{3}, and \WhiteCircled{4}, which are required by Python.
Aside from this, the programmer can directly copy the important logic, e.g., loading the dataset and training the model, from the original implementation.
In addition, if the programmer customizes the model architecture, the model implementation does not require modification.

\noindent{\textbf{Imperative programming interface.}}
This interface does not require the programmer to implement the side task in a step-wise way.
Therefore, instead of implementing the side task in multiple functions (steps \WhiteCircled{2} --- \WhiteCircled{5}), the programmer can merge them in \texttt{RunGpuWorkload()}, as discussed in Section~\ref{sec:template}.
However, this approach trades performance for less programming effort, as pausing side tasks through the framework-enforced mechanism incurs more overheads. 
When the side task manager initiates \texttt{PauseSideTask()} state transition via an RPC at the end of a bubble, even though the CPU process of the side task is paused by the framework-enforced mechanism (Section~\ref{sec:resource-control}) after the state transition, CUDA kernels that have already started cannot be paused because they are asynchronous~\cite{cuda_c}.
As a result, these CUDA kernels will overlap with pipeline training, causing a higher performance overhead than the iterative interface.

\section{Evaluation}
\label{sec:evaluation}

In this section, we evaluate the benefits and overhead of using \bb{} to serve side tasks.

\subsection{Methodology}
\label{sec:meth}
We describe the experimental setup of our evaluation.

\subsubsection{Server setup.}
We use a main server (Server-I) with four RTX 6000 Ada GPUs each with 48 GB of GPU memory to evaluate all pipeline training workloads and side tasks. 
We use a second server (Server-II) with an RTX 3080 GPU with 10 GB of memory to run side tasks separately.
Due to the global shortage of cloud GPUs, we quote prices from a community cloud vendor \cite{runpod_runpod_nodate} that has GPUs available.
The prices of the two servers are $P_{\rm Server-I}=\$3.96\text{/hour}$ and $P_{\rm Server-II}=\$0.18\text{/hour}$, respectively (as of June, 2024). 
The price differences between higher- and lower-tier GPUs in major cloud GPU platforms are similar~\cite{lambdalabs, aws_g4_pricing, aws_p4_pricing}.
In addition to experiments on GPUs, we use a third server (Server-CPU) with 8 cores of an Intel Xeon Platinum 8269Y CPU and 16 GB of memory to evaluate side task performance on CPU. 
We deploy both pipeline training and side tasks in Docker 26.1.2~\cite{bernstein_containers_2014}.

\subsubsection{Comparison points.}\label{subsubsec:comparison_points}
We evaluate \bb{} for side tasks developed with both the iterative and imperative interfaces. 
For comparison, we evaluate MPS \cite{nvidia_developers_multi-process_2024}, where we set pipeline training with the highest priority and side tasks with a lower priority.
We also evaluate a naive co-location approach by directly co-running side tasks and the main pipeline training workload on the same GPU.

\subsubsection{Pipeline training setup.}\label{subsubsec:pipeline_setup}
We train nanoGPT~\cite{karpathy_nanogpt_2024, choi_envpipe_2023} with model sizes 1.2B, 3.6B, and 6B with DeepSpeed 0.12.2~\cite{deepspeed_0_12_2} in a 4-stage pipeline on Server-II (stages 0---3 in Figure~\ref{fig:bubble_study}).
We always maximize the micro-batch size (until just before OOM) to make full use of GPU memory during training.

\subsubsection{Side task workloads.} \label{subsubsec:sidetasks}
We implement 3 types of side tasks: model training, graph analytics, and image processing using both the iterative and the imperative interfaces of \bb{}.
\emph{Model training} side tasks include ResNet18, ResNet50, and VGG19.
We implement the training procedure using out-of-the-box models from PyTorch~\cite{pytorch_developers_models_nodate}.
\emph{Graph analytics} side tasks are adapted from Gardenia~\cite{xu_gardenia_2019}.
It includes PageRank (PR) which is based on the PageRank algorithm~\cite{page_pagerank_1999} and Graph SGD (SGD) which uses stochastic gradient descent to solve matrix factorization~\cite{koren_matrix_2009}, both using the Orkut dataset~\cite{yang_defining_2012}.
The \emph{image processing} (Image) side task resizes an input image and adds a watermark, which we adapt from Nvidia's code~\cite{nvidia_developers_image_2019}.

\subsubsection{Metrics.}

We use \emph{time increase} $I$ and \emph{cost savings} $S$ in Dollars due to side tasks as metrics.
Time increase describes the performance overhead of co-locating side tasks with the main pipeline training workload.
It is the ratio of extra time of pipeline training with side tasks, compared with the original DeepSpeed without any side tasks, and lower time increase means lower overhead.
It is defined as 
\begin{equation*}
    I = \frac{T_{\rm withSideTasks}-T_{\rm noSideTask}}{T_{\rm noSideTask}}~~~.
\end{equation*}

Cost savings describe the benefits of running side tasks.
It is hard to directly measure the benefits of running side tasks for two reasons.
First, the throughput of different size tasks and the main pipeline training workload cannot be directly compared.
Second, the workloads of side tasks are typically deployed on GPUs of smaller scales, while pipeline training mostly uses flagship GPUs.
To compare the value of side tasks and pipeline training that runs on different GPUs with different throughputs and to calculate the benefits, we use their costs (dollars spent on GPUs) as a proxy.
First, we define the cost of pipeline training without side tasks as
\begin{equation*}
    C_{\rm noSideTask} = P_{\rm Server-I} \times T_{\rm noSideTask}
\end{equation*}
and the cost of pipeline training with side tasks as
\begin{equation*}
    C_{\rm withSideTasks} = P_{\rm Server-I} \times T_{\rm withSideTasks}~~~.
\end{equation*}
Then, we compute the cost of running the same side tasks on dedicated lower-tier GPUs as
\begin{equation*}
    C_{\rm sideTasks} = \sum\nolimits_{\rm Each~sideTask} P_{\rm Server-II} \times \frac{W_{\rm sideTask, Server-I}}{\mathit{Th}_{\rm sideTask, Server-II}}~~~
\end{equation*}
where $W_{\rm sideTask, Server-I}$ is the work done by a side task on Server-I, e.g., the number of epochs for model training side tasks, the number of iterations for graph analytics side tasks, and the number of images for the image processing side task.
$\mathit{Th}_{\rm sideTask, Server-II}$ is the throughput of running the same side task on Server-II, which we measure by running side tasks individually on Server-II.
Finally, we define the cost savings $S$ below, where the higher the $S$ value, the greater the benefit.
\begin{equation*}
    S=\frac{C_{\rm sideTasks} - (C_{\rm withSideTasks} - C_{\rm noSideTask})}{C_{\rm noSideTask}}~~~.
\end{equation*}
$C_{\rm sideTasks}$ considers the cost of the side tasks served during bubbles, measured by the cost of running the same side task workloads on dedicated GPUs.
$C_{\rm withSideTasks} - C_{\rm noSideTask}$ is the extra cost from co-location, measured by the increased costs due to co-locating side tasks and pipeline training.

\subsection{Performance Evaluation}
\label{sec:end_to_end}

\begin{figure*}
    \includegraphics[width=\textwidth]{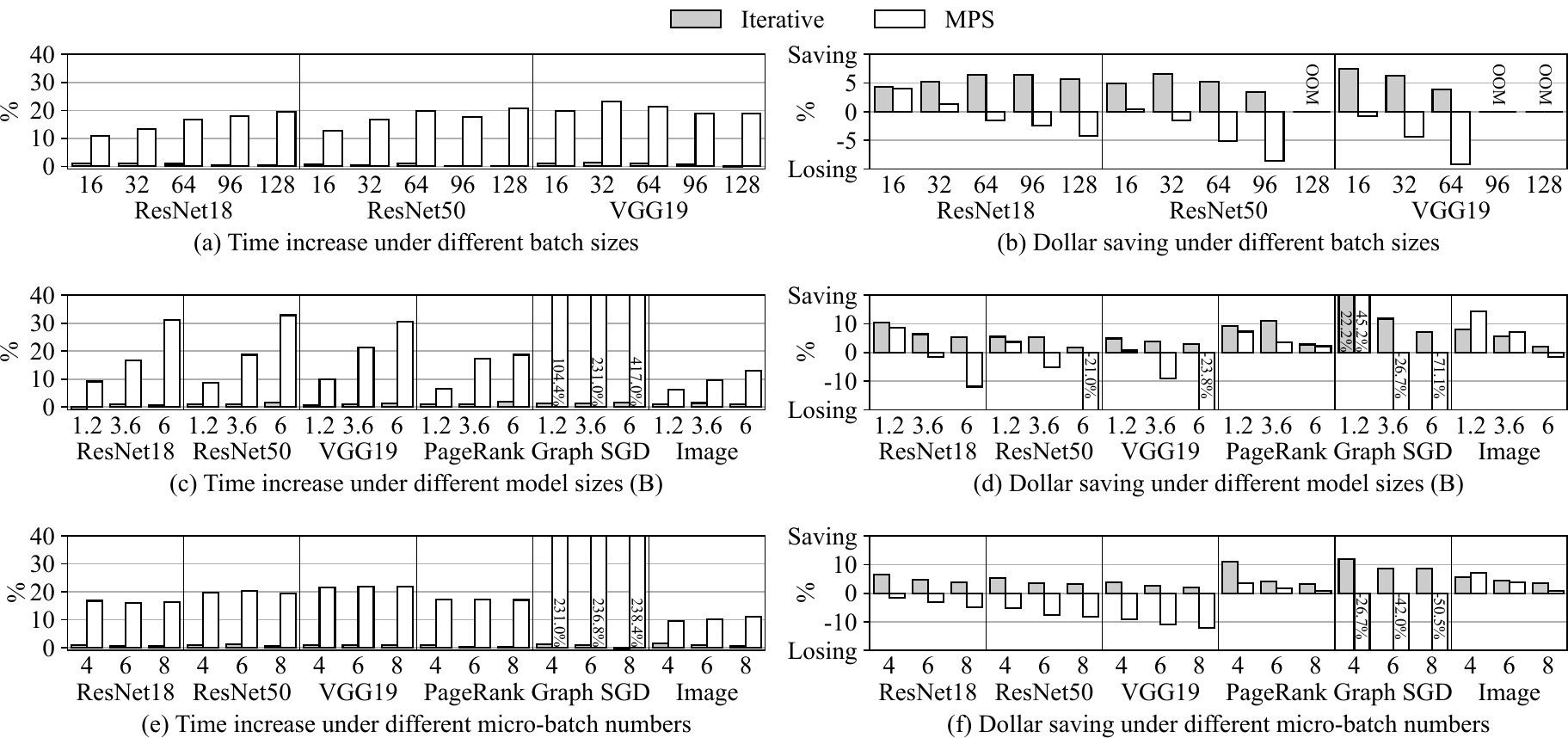}
    \caption{Sensitivity studies of \bb{}.}
    \label{fig:sensitivity}
    \Description[]{}
\end{figure*}

We run DeepSpeed to train a 3.6B model for 128 epochs with side tasks from Section~\ref{subsubsec:sidetasks} and compare the performance overhead, i.e., time increase ($I$) and cost savings ($S$) of using \bb{} with the two interfaces and the two comparative methods (as mentioned in Section~\ref{subsubsec:comparison_points}).
For model training side tasks, we set the batch size as 64.
We run the \emph{same side task} in all workers if they have enough GPU memory. 
We also run a \emph{mixed workload} with 4 side tasks: PageRank, ResNet18, Image, and VGG19, each in one worker corresponding to the GPU of stages 0---3 in Section \ref{subsubsec:pipeline_setup}, respectively.

\begin{table}[t]
\caption{Throughput of GPU side tasks on different platforms measured as iterations per second.}
\label{tab:tput}
\begin{tabular}{cccc}
\toprule
Side task  & Iterative & Server-II & Server-CPU  \\ \midrule
ResNet18  & 1586.6             & 998.7     & 26.5 \\
ResNet50  & 533.1              & 393.4     & 9.1 \\
VGG19     & 170.7              & 161.8     & 3.0 \\
PageRank  & 333.9              & 126.3     & 11.1      \\
Graph SGD & 4.2                & 1.5       & 0.6       \\
Image     & 12.2               & 7.8       & 1.6         \\ \bottomrule
\end{tabular}
\end{table}

\noindent\textbf{Performance compared to lower-tier GPU and CPU.}
Table~\ref{tab:tput} compares the throughput of side task workloads in Section~\ref{subsubsec:sidetasks} running on bubbles using the iterative interface of \bb{} (the Iterative column), on Server-II and Server-CPU as introduced in Section~\ref{sec:meth}.
The throughput is measured as iterations per second.
For ResNet18, ResNet50, and VGG19 GPU workloads, one iteration corresponds to one batch of training data.
For PageRank and Graph SGD, in each iteration, the graph algorithm runs over the input graph for one step.
For image processing workload, one iteration processes one image.

This comparison shows that although side tasks run only during bubbles, they still achieve higher throughput compared to running on a standalone lower-tier GPU (RTX 3080) or the 8-core CPU instance.
By introducing little overhead to pipeline parallel training, \bb{} harvests GPU resources that support a throughput of 1.06--2.82$\times$ of a standalone lower-tier GPU, and 7--59.9$\times$ of the CPU. 
These results demonstrate \bb{}’s effectiveness in harvesting the bubbles in pipeline parallel training.

\begin{table}[t]
\caption{Time increase $I$ (lower the better) and cost savings $S$ (positive=benefit, negative=loss, higher the better) of running DeepSpeed with side tasks.}
\label{tab:e2e}
\setlength{\tabcolsep}{2.5pt}
\resizebox{\linewidth}{!}{
\begin{tabular}{ccccccccccc}
\toprule
 & \multicolumn{4}{c}{\bb{}} & \multicolumn{4}{c}{Comparative Methods} \\
 & \multicolumn{2}{c}{Iterative} & \multicolumn{2}{c}{Imperative} & \multicolumn{2}{c}{Nvidia MPS} & \multicolumn{2}{c}{Naive co-location} \\
Side task   & $I$ \% & ${S}$ \% & $I$ \% & ${S}$ \% & $I$ \% & ${S}$ \% & $I$ \% & ${S}$ \% \\
\midrule
ResNet18    & 0.9   & 6.4   & 2.2   & 6.0   & 16.8  & -1.5  & 49.8  & -30.7 \\
ResNet50    & 0.9   & 5.3   & 3.8   & 3.9   & 19.8  & -5.1  & 61.9  & -44.0 \\
VGG19       & 0.9   & 3.9   & 5.0   & 1.4   & 21.4  & -9.1  & 53.4  & -39.7 \\
PageRank    & 1.0   & 11.1  & 2.5   & 16.4  & 17.3  & 3.5  & 45.1  & -16.0 \\
Graph SGD   & 1.2   & 11.8  & 4.1   & 22.8  & 231.0 & -26.7 & 62.4  & -9.1  \\
Image       & 1.4   & 5.7   & 2.7   & 6.1   & 9.5   & 7.2   & 46.0  & -29.3 \\
Mixed       & 1.1   & 10.1  & 4.3   & 11.0  & 24.8  & 0.2   & 64.3  & -35.5 \\
\bottomrule
\end{tabular}
}
\end{table}

\noindent\textbf{Co-location performance.}
Table~\ref{tab:e2e} shows the time increase and cost savings of running DeepSpeed with side tasks of different co-location methods.
\bb{} consistently exhibits lower overhead than the comparative methods, showing only a 1.1\% average time increase while achieving 7.8\% average cost savings through side tasks using the iterative interface.
The imperative interface achieves comparable cost savings but with a higher overhead as it relies on the less efficient framework-enforced mechanism to limit the side task's execution time (Section~\ref{sec:resource-control}).
In comparison, the average time increase and cost savings for MPS are 48.7\% and -4.5\%, and for Naive are 54.7\% and -29.2\%.
Their \emph{negative} cost savings indicate that these approaches can increase the total cost. 
Notably, the time increase of Graph SGD with MPS is as high as 231.0\%.
This anomaly is due to Graph SGD's high compute intensity.
We conclude that \bb{} effectively utilizes bubbles in pipeline training for serving side tasks.
While the comparative methods can utilize bubbles, unlike \bb{}, they are not designed for this purpose. Thus, they are inefficient in using bubbles, leading to higher costs.

\subsection{Sensitivity Study}
\label{sec:sensitivity}

We change the side task batch size, DeepSpeed model size, and DeepSpeed micro-batch numbers of side tasks, and study the time increase and cost savings of \bb{} with the iterative interface.

\noindent\textbf{(1) Varying batch sizes.}
Figure~\ref{fig:sensitivity}(a) includes model training side tasks under variable batch sizes.
Other side tasks are not included as they do not run with batch sizes.
\textit{OOM} means that the GPU in Server-II does not have enough GPU memory for the configuration, so the cost savings cannot be calculated.
\bb{} has low performance overheads, with around 1\% increase in execution time, and cost savings of 3.4\% -- 7.5\%.

\noindent\textbf{(2) Varying model sizes.}
In Figure~\ref{fig:sensitivity}(b), the performance overheads of \bb{} range from -0.7\% to 1.9\%, and cost savings range from 1.8\% to 22.2\%.
The main reason is the shorter bubble durations when training larger models as the main workload, which was also shown in Figure~\ref{fig:duration-memory}.

\noindent\textbf{(3) Varying micro-batch numbers.}
In Figure~\ref{fig:sensitivity}(c), the performance overhead of \bb{} increases from -0.4\% to 1.5\%, and cost savings reduces from 2.1\% to 11.8\%.
When the micro-batch number increases, because of the lower bubble rate (Section~\ref{sec:bubble-study}), the cost savings decrease.

\begin{figure}
    \centering
    \includegraphics[width=\columnwidth]{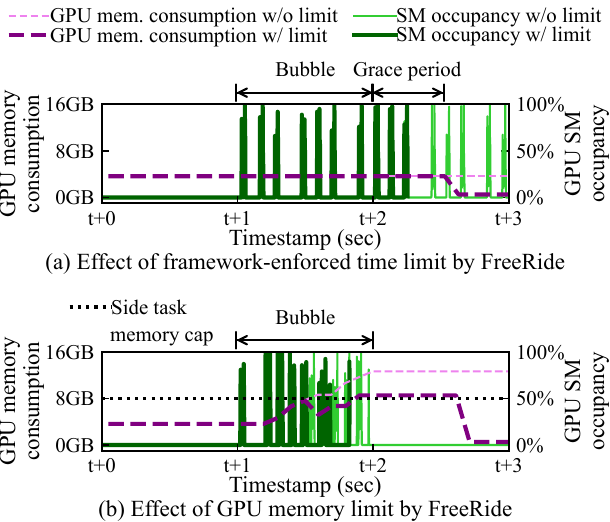}
    \caption{Demonstration of GPU resource limit in \bb{}.}
    \label{fig:constraint}
    \Description[]{}
\end{figure}

\subsection{Effectiveness of GPU Resource Limit}
\label{sec:resource_limit_exp}

We use training ResNet18 as an example to demonstrate the GPU resource limit mechanism in \bb{}.

\noindent\textbf{Side task execution time limit.}
Figure~\ref{fig:constraint}(a) demonstrates a case where the side task does not pause after the bubble that ends at $t+2$.
With GPU resource limit, as shown by the green and purple curves, the worker terminates the side task after a grace period via the framework-enforced mechanism.

\noindent\textbf{Side task GPU memory limit.}
Figure~\ref{fig:constraint}(b) illustrates another case where the side task keeps allocating GPU memory despite its 8 GB limit. 
Without \bb{}'s GPU resource limit mechanism, 
the side task's GPU memory allocation is only capped by the physical memory limit of the GPU, potentially interfering with the main training workload.
With GPU resource limit, after the side task process exceeds its 8 GB GPU memory limit, it is terminated to release GPU memory.

\subsection{Bubble Time Breakdown}
\label{sec:bubble_efficiency}

\begin{figure}
    \centering
    \includegraphics[width=\columnwidth]{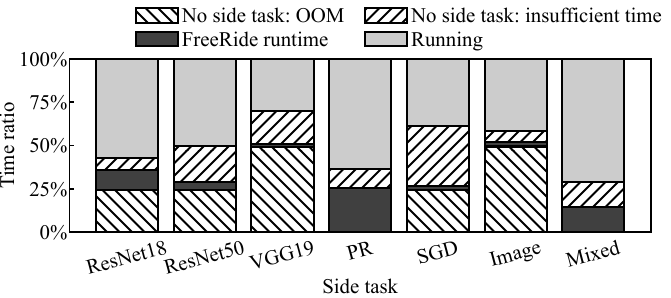}
    \caption{Bubble time breakdown.}
    \label{fig:bubble_breakdown}
    \Description[]{}
\end{figure}

In Figure~\ref{fig:bubble_breakdown}, we present a breakdown of bubble utilization in \bb{} under the iterative interface.
\textit{No side task: OOM} means that some bubbles are unused due to their limited available GPU memory.
For instance, the GPU memory consumption of VGG19 or the Image side task is larger than the GPU memory of bubbles in stages 0 and 1, so they cannot use half of the bubble time.
\textit{No side task: insufficient time} refers to idle time because the remaining time of a bubble is not enough for the next step of the side task.
\textit{\bb{} runtime} is the time consumed by running \bb{}, including the interface code and the side task manager.
Most of the bubble time with enough available GPU memory size is used by side tasks.
For side tasks with shorter per-step durations, e.g., PageRank, the ratio of \bb{} runtime is higher because more iterations of the iterative interface are executed.
In contrast, side tasks with longer per-step durations have lower bubble utilization because of insufficient time. 

\section{Related Work}
\label{sec:related}

\noindent\textbf{Pipeline parallelism and bubbles.}
Prior work has aimed to improve the schedule of pipeline training to reduce bubbles~\cite{rasley_deepspeed_2020, deepspeed_developers_pipeline_2023, kim_torchgpipe_2020, huang_gpipe_2019, fan_dapple_2021, liu_hanayo_2023, li_chimera_2021, harlap_pipedream_2018, narayanan_memory-efficient_2021, tang_adaptive_2024, qi_zero_2023}.
Other work leverages bubbles in \pp{} by assigning specialized procedures coupled with pipeline training to enable fault tolerance through replicated computation~\cite{thorpe_bamboo_2023}, or for accelerating the training algorithms~\cite{osawa_pipefisher_2023, hydro}.
These approaches require changes to the training framework and are limited to certain types of workloads.
In contrast, \bb{} does not require any changes to, or coupling with, pipeline training to serve generic GPU side tasks.

\noindent\textbf{GPU sharing.}
Gandiva time-slices GPUs for multiple jobs with fallback to non-sharing GPUs~\cite{xiao_gandiva_2018}.
Salus designs job switching and memory sharing primitives for GPU sharing~\cite{yu_salus_2019}.
PipeSwitch further improves GPU sharing by designing fast context switch mechanisms between the host memory and the GPU memory~\cite{bai_pipeswitch_2020}.
AntMan designs dynamic scaling mechanisms for distributed deep learning workloads~\cite{xiao_antman_2020}.
Zico tracks the GPU memory allocation and reclamation of deep learning jobs and shares the GPU memory reclaimed by one job with other jobs~\cite{lim_2021_zico}.
Veltair proposes a compiler that co-optimizes the compiling results of co-located GPU workloads~\cite{veltair}.
TGS achieves GPU sharing of deep learning workloads in container clouds through rate control and transparent shared memory at the OS level~\cite{wu_transparent_2023}.
Recently, Orion proposes GPU sharing by intercepting and scheduling CUDA kernel calls made by PyTorch~\cite{strati_orion_2024}.
While these approaches propose methods to share GPUs and continuously improve the utilization of GPUs, they share the GPUs aggressively, without minimizing the impact on high-priority and high-cost workloads.
Therefore, they would cause high overhead (time increase) if used to co-locate LLM training and side tasks, and subsequently yield little to no cost savings due to their high overhead.
In comparison, FreeRide achieves GPU sharing while maintaining a very low overhead.
PilotFish on the other hand leverages the free cycles in cloud gaming for deep learning workloads~\cite{zhang_pilotfish_2022}, while FreeRide harvests the bubbles in pipeline parallelism for other generic GPU workloads.
MPS and MIG~\cite{nvidia_developers_multi-process_2024, nvidia_developers_mig} are mechanisms provided by Nvidia for GPU sharing and virtualization.
\bb{} leverages MPS to impose GPU memory limits on side tasks.

\section{Discussion}\label{sec:discussion}

\noindent{\textbf{Security.}}
Prior GPU sharing solutions tend to prioritize efficiency and assume a safe environment~\cite{yu_salus_2019, xiao_gandiva_2018, zhang_pilotfish_2022, lim_2021_zico, han_2022_reef}.
E.g., Orion assumes that the co-located GPU workloads are in the same trust domain~\cite{strati_orion_2024}.
\bb{} provides the same security and isolation guarantees as the lower-level system it is built on.
It incorporates MPS to limit GPU memory which provides separate GPU address spaces~\cite{mps_memory_protection} for pipeline training and side tasks, and Docker for environment isolation~\cite{docker_security, yasrab_2023_mitigating}.
Orthogonally, security for co-located GPU workloads is an active research area~\cite{pavlidakis_2024_gsafe, Naghibijouybari_2018_rendered, Naghibijouybari_2021_side, kim_2020_gpu, liu_2019_side, zhang_2024_bridge}.
We expect future work to co-design security with efficient GPU sharing. 

\noindent{\textbf{Fault tolerance.}}
Since \bb{} supports generic side tasks, it is not possible for \bb{} to exhaustively implement fault tolerance mechanisms for all side tasks.
Instead, \bb{} assumes that side tasks implement their own fault tolerance mechanisms to tolerate the failure of side tasks themselves and of pipeline training.
In addition, since \bb{} deploys side tasks in Docker containers as processes that are independent of the pipeline training, failures of side tasks, such as illegal memory access, will not impact the main pipeline training workload.

\noindent{\textbf{Side task management.}}
By implementing different strategies in its side task manager, \bb{} can incorporate more sophisticated management, e.g., co-locating multiple side tasks with various performance characteristics in the same worker to improve the utilization of bubbles~\cite{veltair} or serving side tasks with fairness or performance guarantees~\cite{drf, chen_prophet_2017}.

\noindent{\textbf{Scalability.}}
\bb{} can be extended for better scalability.
As \bb{} implements communication among its components using RPCs, it can be easily extended to distributed settings with side tasks on multiple servers.
During training, the side task manager of \bb{} receives bubbles from all GPUs from both remote servers and manages the side tasks that co-locate with each GPU.
\bb{} can also be extended to support multi-GPU side tasks, e.g., distributed training and big data processing~\cite{mapreduce, ghive_socc}, by launching workers with access to multiple GPUs.

\noindent{\textbf{Stability of pipeline training.}}
\bb{} follows the same assumption as the previous pipeline parallel training works that pipeline training has a stable throughput and pattern, and that the training sequences have the same length after padding~\cite{liu_hanayo_2023,fan_dapple_2021,huang_gpipe_2019,qi_zero_2023}.

\noindent{\textbf{Other ML accelerators.}}
This work targets GPUs due to their widespread accessibility.
\bb{}'s mitigation for bubbles fundamentally applies to other ML accelerators, such as Google's TPU~\cite{tpu2017isca} and Meta's MTIA~\cite{mtiav1}, provided that the platform has isolation and resource limit options for each process.
We anticipate future work to incorporate the approach of \bb{} with other ML platforms. 

\noindent\textbf{Energy consumption.}
There has been recent interest in building energy- and carbon-efficient systems for machine learning workloads~\cite{choi_envpipe_2023, ecoserve, greenflow, stojkovic2024dynamollm, samsi2023words, nguyen2024towards}.
We anticipate future work on energy efficiency of co-locating side tasks with LLM training.

\section{Conclusion}
\label{sec:conclusion}
We propose \bb{}, a middleware system that bridges the gap between the available yet hard-to-utilize bubbles in pipeline parallelism and running generic GPU side tasks to harvest them.
It provides programming interfaces that abstract the life cycle of a side task as different states of a state machine and allows programmers to implement side tasks with little engineering effort.
The side task manager and side task workers manage bubbles and side tasks, and reduce the performance overhead of side tasks on pipeline training.
Our evaluation shows that, on average, \bb{} achieves 7.8\% cost savings for long-running and expensive pipeline training with a negligible performance overhead of about 1\%.

\begin{acks}
We thank the anonymous reviewers and our shepherd Ruben Mayer for their constructive feedback to improve this paper.
This work was supported by the Natural Sciences and Engineering Research Council of Canada (NSERC) and the Undergraduate Research Assistantship program of the Cheriton School of Computer
Science at the University of Waterloo.

\end{acks}

\balance

\bibliographystyle{plain}
\bibliography{ref}

\end{document}